\documentclass[twocolumn,prX,nofootinbib]{revtex4-2}
\usepackage{amsmath,amsfonts,amssymb,amsthm,array,eucal,mathrsfs,upgreek,stackrel,slashed,graphicx,tensor,color,enumerate,mathtools,textcomp,verbatim,caption,hyperref}

\usepackage{slashed}
\usepackage{braket}
\usepackage{bm}
\usepackage{dsfont}

\DeclareMathOperator{\hil}{\mathcal{H}}
\DeclareMathOperator{\Tr}{\text{Tr}}
\DeclareMathOperator{\erf}{\text{erf}}

\newcommand{\beq}{\begin{equation}}
\newcommand{\beql}[1]{\begin{equation}\label{#1}}
\newcommand{\eeq}{\end{equation}}
\newcommand{\eeqp}{\;\;\;.\end{equation}}
\newcommand{\eeqc}{\;\;\;,\end{equation}}

\begin{document} 
\title{Energy Cost of Localization of Relational Quantum Information}

% authors and affiliation
\author{A. Dukehart}\email{Adam.Dukehart@unh.edu}
\author{and D. Mattingly}\email{David.Mattingly@unh.edu}
\affiliation{University of New Hampshire,\\Durham, NH, USA}

% e-mail addresses

\begin{abstract}
Entanglement of spatially separated quantum states is usually defined with respect to a reference frame provided by some external observer. Thus, if one wishes to localize the quantum information within a spatially separated entangled state, one must enact an entanglement extraction protocol also defined with respect to that external frame.  Entanglement extraction for Gaussian ground states in such an external frame construction has been shown to require a minimum energy and is hence an interesting process for gravitational physics, where examinations of localization vs. energy cost have a long history.   General covariance however, precludes dependence on external frames. In order to enact an extraction protocol in a generally covariant theory, dependence on the external reference frame must first be removed and the states made relational. We examine the implementation of an extraction protocol for Gaussian states, who's center-of-mass and relational degrees of freedom are entangled, in a relational toy model where translation invariance stands in for full diffeomorphism invariance.  Constructing fully relational states and the corresponding extraction/localization can, in principle, be done in two ways.  External frame position information can be removed through $G$-twirling over translations or one can spontaneously break the translation symmetry via the gradient of an auxiliary field, or $Z$-model.   We determine the energetics of quantum information localization after the states have been made fully relational via both the $G$-twirl and $Z$-model. We also show one can smoothly transition between the two approaches via positive operator valued measurements (POVM).
\end{abstract}

\maketitle
%\flushbottom

%---------------------------------------------------------------------------------------------------------%
%---------------------------------------------- INTRODUCTION ---------------------------------------------%
%---------------------------------------------------------------------------------------------------------%
\section{Introduction}
In quantum gravity phenomenology, modifications to the physics of the standard model plus general relativity often come with a length scale $L$ attached~\cite{Hossenfelder2013}.  This scale, perhaps in conjunction with a second scale in a hierarchy or an experimental length scale, suppresses the new phenomenology so that the effect is either negligible or almost negligible in our currently achievable experiments.  The exact size of $L$ can come from many places\textemdash it may be the string scale~\cite{KonishiPaffutiProvero1990}, the scale at which the unitarity of the standard model plus general relativity (treated as an effective field theory) breaks down~\cite{HanWillenbrock2005}, or some postulated scale where there is new, exotic spacetime structure such as non-commutative spacetime~\cite{HinchliffeKerstingMa2004}. In many scenarios, the scale is typically at or near the Planck scale.

Independent of the quantum gravity model underlying the choice of scale, the Planck scale is often singled out heuristically by a simple (albeit somewhat flawed) argument.  A massless particle trapped in a box of size $L$ has roughly an energy of $E(L)=h c/L$ in the rest frame of the box, neglecting dimensionless constants.  As $L$ shrinks, the energy goes up and eventually passes the energy necessary for a black hole to form in that region. i.e. if $R(E)$ is the Schwarszchild radius for an energy $E$, $R(E)=2GE/c^4$, then when $L\approx R(E(L))$ black holes will begin to form, rapid Hawking emission will occur, and one is in the quantum gravity regime.  This occurs at $L\approx \sqrt{Gh/c^3}$, \emph{i.e.} somewhere around the Planck length.

The above argument can be viewed as the breakdown of locality: one cannot measure distances below the Planck scale because the act of doing so would disturb spacetime to the degree that we would form black holes. There is an energy cost to localizing a particle, and this energy eventually backreacts on the spacetime~\footnote{See e.g.~\cite{Banks2003} for more formal arguments on high energy physics being dominated by black hole states, or asymptotic darkness.}.  This construction, however, generally assumes the particle is in an energy eigenstate of some external observer and localized to a particular region.  In other words, the state in the most naive version is fundamentally assumed to be in a product state: $\ket{\Psi}=\ket{\Psi_{sys}} \otimes \ket{\Psi_{env}}$, where $\ket{\Psi_{sys}}$, the particle, can be tuned independently.  

Quantum mechanics does not, of course, have only product states in any given Hilbert space\textemdash one can have entangled states as well.  In this case, the information contained in the state can be non-local due to the entanglement in addition to any non-locality inherent in the underlying basis states.  Since we have many examples of black hole physics and geometric surfaces being re-understood in terms of quantum information processes (e.g.~\cite{Harlow2017} among many others.), and since the above argument on the breakdown of locality critically relies on black hole formation, a natural question is then to ask about the energetics of localizing quantum information in entangled states and the possible gravitational effects.

Whether or not quantum information can be arbitrarily localized in gravity is currently a matter of some debate. In the classical picture, due to diffeomorphism invariance, energy and momentum cannot be localized and one must resort to a quasilocal picture. In perturbative quantum gravity, Giddings and Donnelly have argued that at least at first order in Newton's constant $G_N$, quantum information can indeed be localized~\cite{DonnellyGiddings2018}. However, there are also arguments stemming from holography, in particular boundary unitarity, that indicate that a purely localized operator in the bulk would be inconsistent~\cite{JacobsonNguyen2019}. The question also comes into play as arguments about the energetics of entanglement entropy in local regions have also been used to derive general relativity from more fundamental principles~\cite{Jacobson2016}.

In this paper we examine information localization and energy costs from a quantum information perspective using non-local quantum information contained in bipartite entangled Gaussian states.  In this approach, the question of localization can be recast as one of entanglement extraction, where extracting the entanglement corresponds to localizing the system.  Hackl and Jonsson have recently made progress on calculating the minimum energy necessary to extract such bipartite entanglement from Gaussian systems, which will provide the tools necessary to derive the energetics for our quantum information localization process~\cite{HacklJonsson2019}.  

There is a complication, however.  Besides the issue of entanglement, the standard naive argument relied on the existence of an external observer to set the energy scale and frame.  In general relativity of course there are no external observers and the observables are expected to be relational.  In order to apply Hackl and Jonsson's approach to a generally covariant theory, one must also construct a system where the degrees of freedom are relational and Gaussian.  If not, then one would not be sure whether the answers are gauge invariant and hence physical.  This can be done in our toy model in two ways. First, one can apply a procedure known as $G$-twirling to a set of Gaussian states in the presence of an external frame.  By $G$-twirling over translations, we remove the notion of absolute position in our system, leaving only relational and entangled positional degrees of freedom~\cite{SmithPianiMann2016} (as well as an irrelevant center-of-mass momentum degree of freedom).  As we shall see, the simplest version of $G$-twirling implies that the energy cost from entanglement extraction vanishes.  This is consistent with the implementation of the $G$-twirl as a simple transformation on the Hilbert space. The underlying translation invariance and effect on the energy cost is a toy model equivalent of the effect of diffeomorphism invariance and the vanishing of the Hamiltonian on physical states in quantum gravity.

In contrast, one can also implement an external frame as a dynamical system, thereby also relationalizing the quantum states. We will show that in this case we naturally recover a non-zero energy cost.  To implement this frame, we introduce a $U(1)$ symmetry and corresponding gauge field, and build a simple $Z$-model for relational observables in the language of Giddings, Marolf, and Hartle~\cite{GiddingsMarolfHartle2006}.  

The paper is constructed as follows.  In Section ~\ref{sec:GaussianStates} we outline the fundamental construction of $N$ particle Gaussian states centered at different positions in the presence of an external partition.  In Section~\ref{sec:Fundamentals} we introduce the different fundamental techniques we will employ for the calculation.  In Section~\ref{sec:RelationalSystems}, we apply these various techniques specifically analyze the question of entanglement extraction in different relational constructions. Finally, we conclude in Section~\ref{sec:Conclusion}. Throughout this paper we work in $\hbar = m = 1$ units.
%\\
%----------------------\\

\section{Gaussian states with external partitions}\label{sec:GaussianStates}
\subsection{Correlation functions and Gaussian states}\label{sec:Correlation}
Since we will be working with Gaussian states, we first provide some background on the mathematics of Gaussian states and their entanglement. Generally, Gaussian states provide a versatile analytical tool in many areas of quantum physics, although we will only use a small subset of their power herein.  Unless explicitly noted we work in one spatial dimension.  We also follow the presentation in~\cite{HacklBianchi2021}, which allows for the treatment of bosons and fermions in a unified framework.  
 
 The standard approach assumes the ability to establish a canonical $p_i,q_i$ basis of an $N$ particle phase space, where $i$ runs from $1$ to $N$.  As such, it implicitly uses an external partition\textemdash the division of the overall phase space into the subspaces associated with each particle and the labeling of position and momentum as measured by some external measurement system.  In the language of quantum reference frames, such an external partition is called a ``perfect'' quantum reference frame~\cite{SmithPianiMann2016,BartlettRudolphSpekkens2007}.  In such a frame each single particle Hilbert space is spanned by some continuous set of kets $\ket{g}$ that are completely distinguishable. In this case the typical choice is the position basis kets $\ket{x}$, with distinguishability implemented as $\braket{x|x'}=\delta(x'-x)$ (or equivalently the momentum basis kets).  This distinguishability is then reflected in the classical $p,q$ phase space in some associated classical reference frame.  For now, we will utilize such a frame both quantum mechanically and classically, although we return to this point later.

With such a frame, for a system with $N$ particle degrees of freedom, we have the classical phase space $V\simeq\mathbb{R}^{2N}$, consisting of the $p$'s and $q$'s from above, and its dual $V^*\simeq\mathbb{R}^{2N}$.  We quantize the system by promoting the $2N$ phase space coordinates to operators representing observables, that can be put in an operator valued vector $\hat{\xi}^a=(\hat{q}_1,\hat{q}_2,...,\hat{q}_N, \hat{p}_1,\hat{p}_2,...,\hat{p}_N)$.  A Gaussian state $\ket{\psi}$, whether it is bosonic or fermionic, is completely described by the one-point\footnote{Note that $z^a = 0$ for fermions.}, $z^a := \bra{\psi}\hat{\xi}^a\ket{\psi}$ and two-point $C^{ab}_2 := \bra{\psi} (\hat{\xi}-z)^a (\hat{\xi}-z)^b \ket{\psi}$, correlation functions (\emph{cf}.~\cite{FerraroOlivaresParis2005}). All higher order correlations can be determined from the one- and two-point correlation functions. The two-point correlation function can be decomposed into a symmetric piece $G^{ab}$ and an antisymmetric piece $\Omega^{ab}$ via
\begin{equation}
    C^{ab}_2 = \frac{1}{2}\left(G^{ab} + i \Omega^{ab}\right).
\end{equation}

\subsection{Bosonic vs. fermionic degrees of freedom}\label{sec:DOF}
For bosonic and fermionic degrees of freedom, the roles and behavior of $G^{ab}$ and $\Omega^{ab}$ differ. Bosonic degrees of freedom are characterized by the commutation relations
\begin{eqnarray}
    \left[\hat{x}^i,\hat{p}^j\right] &=& i\hbar \delta^{ij} \hat{\mathbf{I}},\\
    \left[\hat{x}^i,\hat{x}^j\right] = [\hat{p}^i,\hat{p}^j] &=& 0.
\end{eqnarray}
We can isolate the antisymmetric $\Omega^{ab}$ via
\begin{equation}\label{eqn:Omega}
    \bra{\psi}\left(\hat{\xi}^a \hat{\xi}^b -\hat{\xi}^b \hat{\xi}^a\right)\ket{\psi}= i \Omega^{ab}
\end{equation}
and expressing $\hat{\xi}^a$ in terms of the phase space operators shows that $\Omega^{ab}$ is simply the symplectic form inherited from the classical Poisson brackets after quantization.  In other words, for bosons
\begin{equation}
    \Omega^{ab}= \left(\begin{tabular}{cc}
        0 & $-i\mathds{I}$\\
        $i\mathds{I}$ & 0
    \end{tabular}\right)
\end{equation}
and is not state dependent. On the other hand, 
\begin{equation}\label{eqn:G}
    \bra{\psi}\left(\hat{\xi}^a \hat{\xi}^b +\hat{\xi}^b \hat{\xi}^a\right)\ket{\psi}= G^{ab}
\end{equation}
shows that $G^{ab}$ is a state dependent quantity for bosons.  This dependence is one-to-one, \emph{i.e.}, any bosonic Gaussian state can be uniquely specified (up to a phase) by $z^a$ and $G^{ab}$.

Conversely, for fermionic degrees of freedom the (anti)-commutation relations are
\begin{eqnarray}
    \{\hat{x}^i,\hat{x}^j\} = \{\hat{p}^i,\hat{p}^j\} &=& \hbar\delta^{ij}\hat{\mathbf{I}},\\
    \{\hat{x}^i,\hat{p}^j\} &=& 0.
\end{eqnarray}
For fermionic degrees of freedom, the roles and behavior of $G^{ab}$ and $\Omega^{ab}$ are reversed, but determined in the same manner as their bosonic counterparts. The symmetric $G^{ab}$ can be isolated via (\ref{eqn:G}). Given the (anti)-commuting nature of fermions, it is clear that $G^{ab}$ is the symmetric, positive-definite, bilinear form inherited from the classical (anti)-commuting Poisson brackets after quantization. This implies that $G^{ab}$ takes the form,
\begin{equation}\label{eqn:G-Form}
    G^{ab} = \left(\begin{tabular}{cc}
        $\mathds{I}$ & 0 \\
        0 & $\mathds{I}$
    \end{tabular}\right)
\end{equation}
and is state independent. The antisymmetric form $\Omega^{ab}$ is determined via (\ref{eqn:Omega}), however for fermions $\Omega^{ab}$ is now a state dependent quantity and is in one-to-one correspondence with each Gaussian state up to a phase.

\subsection{Combined K{\"a}hler structure}\label{sec:Kahler}
While we will concentrate primarily on fermions, it will be useful to do so in a notation that allows for both bosonic and fermionic analysis.  This can be accomplished by unifying the mathematical description of bosonic and fermionic Gaussian states via K{\"a}hler structures. A K{\"a}hler space is a real vector space that is equipped with the following linear operators:
\begin{itemize}
    \item {\bf Metric}, a symmetric, positive-definite, bilinear form $G^{ab}$, with inverse $G_{ab}^{-1}$ such that $G^{ac}G^{-1}_{cb} = \delta^a_b$,
    \item {\bf Symplectic form}, an antisymmetric, non-degenerate form $\Omega^{ab}$, with inverse $\Omega^{-1}_{ab}$ such that $\Omega^{ac}\Omega^{-1}_{cb} = \delta^a_b$,
    \item {\bf Complex structure}, denoted $J^a_b$, satisfies the property $J^a_cJ^c_b = -\delta^a_b$. 
\end{itemize}
The triple of these three operators $(G,\Omega,J)$ is referred to as a K{\"a}hler structure. The three operators are related via,
\begin{equation}\label{eqn:J}
    J^a_b = -G^{ac}\Omega^{-1}_{cb} = \Omega^{ac}G^{-1}_{cb}.
\end{equation}

It is apparent that the bosonic and fermionic Gaussian state spaces have two of the three required linear operators for a K{\"a}hler structure, particularly a metric $G^{ab}$ and a symplectic form $\Omega^{ab}$. However, this is not enough to imply that $G^{ab}$ and $\Omega^{ab}$ are compatible K{\"a}hler structures. We must require that $J^a_b$ defined by $G^{ab}$ and $\Omega^{ab}$ via (\ref{eqn:J}) satisfies the condition $J^2 = -\mathds{I}$ for $G^{ab}$ and $\Omega^{ab}$ to be compatible K{\"a}hler structures.

Bosonic Gaussian states have an associated metric $G^{ab}$ that is state dependent and a symplectic form $\Omega^{ab}$ that is state independent. Assuming both $G^{ab}$ and $\Omega^{ab}$ are K{\"a}hler compatible, the complex structure $J^a_b$, that relates $G^{ab}$ and $\Omega^{ab}$, is therefore a unique state dependent quantity. Similarly, fermionic Gaussian states have a metric $G^{ab}$ that is state independent and a symplectic form $\Omega^{ab}$ that is state dependent. Again, assuming $G^{ab}$ and $\Omega^{ab}$ are K{\"a}hler compatible, the complex structure that relates the two is also state dependent in a similar way. Thus, for both bosonic and fermionic Gaussian states, the complex structure $J^a_b$ is uniquely determined by the state, up to a phase and we can use $J^a_b$ as an ideal label for either bosonic or fermionic Gaussian states.

Furthermore, $J^a_b$ can be used to explicitly define an operator that annihilates the associated Gaussian state.  Given a K{\"a}hler structure, every Gaussian state $\ket{\psi}$ associated with the specified structure solves the equation
\begin{equation}\label{eqn:gausscondition}
    \frac{1}{2}\left(\delta^a_b + iJ^a_b\right)\left(\xi - z\right)^b\ket{\psi} = 0.
\end{equation}

\subsection{Annihilation and creation operators}\label{sec:Ops}
While Gaussian states can be described using sets of phase space operators $\hat{p}_i,\hat{q}_i$ it is also convenient to describe Gaussian states using the Fock basis construction and creation/annihilation operators.

As is familiar from introductory quantum mechanics, a Hilbert space representation of the algebra of observables in the Fock basis can be defined by a set of annihilation and creation operators $\hat{a}_i,\hat{a}^{\dagger}_i$, where $i=1\hdots N$ for a system with $N$ particles. For a system of bosonic particles, we impose the canonical commutation relations on the annihilation and creation operators,
\begin{eqnarray}
    \left[\hat{a}_i,\hat{a}^{\dagger}_j\right] &=& \delta_{ij}\hat{\mathbf{I}},\\
    \left[\hat{a}_i,\hat{a}_j\right] = \left[\hat{a}^{\dagger}_i,\hat{a}^{\dagger}_j\right] &=& 0.
\end{eqnarray}
For a system of fermionic particles, we impose (anti)-commutation relations on the annihilation and creation operators,
\begin{eqnarray}
    \left\{\hat{a}_i,\hat{a}^{\dagger}_j\right\} &=& \delta_{ij}\hat{\mathbf{I}},\\
    \left\{\hat{a}_i,\hat{a}_j\right\} = \left\{\hat{a}^{\dagger}_i,\hat{a}^{\dagger}_j\right\} &=& 0. 
\end{eqnarray}
For both bosons and fermions the vacuum state $\ket{0,\hdots,0}$ is the state annihilated by all $\hat{a}_i$,
\begin{equation}
    \hat{a}_i\ket{0,\hdots,0}=0.
\end{equation}

Orthonormal basis states are given by $\ket{n_1,\hdots,n_N}$ where $n_i\in\mathds{N}$ for bosonic systems and $n_i\in\{0,1\}$ for fermionic systems. The action of the annihilation and creation operators on these states satisfies,
\begin{eqnarray}
    \hat{a}_i\ket{n_1,\hdots,n_N} &=& \sqrt{n_i}\ket{n_1,\hdots,n_i-1,\hdots,n_N},\\
    \hat{a}^{\dagger}_i\ket{n_1,\hdots,n_N} &=& \sqrt{n_i+1}\ket{n_1,\hdots,n_i+1,\hdots,n_N}
\end{eqnarray}
and they can be obtained from the vacuum state via
\begin{equation}
    \ket{n_1,\hdots,n_N} = \prod_{i=1}^N\left(\frac{(\hat{a}^{\dagger}_i)^{n_i}}{\sqrt{n_i!}}\right)\ket{0,\hdots,0}.
\end{equation}

To relate the annihilation and creation operators to the operator valued vector $\hat{\xi}^a$ we need to define transformations $v_{ia}\in V^*_{\mathds{C}}$ in a complex vector space $V^*_{\mathds{C}}$ such that, for a $\hat{\xi}^a$ in some basis,
\begin{eqnarray}
    \hat{a}_i = v_{ia}\hat{\xi}^a,\\
    \hat{a}^{\dagger}_i = v^*_{ia}\hat{\xi}^a.
\end{eqnarray}
Given the relation between $G^{ab}, \Omega^{ab}$ and $\xi^a$ from (\ref{eqn:G}) and (\ref{eqn:Omega}), respectively, it is easy to see that the $\hat{\xi}^a$ associated with bosonic systems inherits the commutation relations and the (anti)-commutation relations for fermionic systems. Along with the algebras for bosonic and fermionic annihilation and creation operators, this implies that there are conditions that the transformations $v_{ia}$ must satisfy. For bosons, the $v_{ia}$ must satisfy,
\begin{eqnarray}
    \Omega^{ab}v_{ia}v_{jb} &=& 0,\\
    \Omega^{ab}v^*_{ia}v_{jb} &=& i\delta_{ij}.
\end{eqnarray}
And similarly for fermions the $v_{ia}$ must satisfy,
\begin{eqnarray}
    G^{ab}v_{ia}v_{jb} = 0,\\
    G^{ab}v^*_{ia}v_{jb} = \delta_{ij}.
\end{eqnarray}
Using these conditions we can define a set of vectors $u^a_j$, dual to $v_{ia}$, that can be used to define a basis transformation between the Fock basis $(\hat{a}_1,\hat{a}^{\dagger}_1,\hdots,\hat{a}_N,\hat{a}^{\dagger}_N)$ and $\hat{\xi}^a$. The vectors $u^a_i$ are defined by
\begin{equation}
    u^a_i = i\Omega^{ab}v^*_{ib},
\end{equation}
for bosons and 
\begin{equation}
    u^a_i = G^{ab}v^*_{ib},    
\end{equation}
for fermions. Given a set of $u^a_i$ the transformation between the Fock basis and $\hat{\xi}^a$ is
\begin{equation}
    \hat{\xi}^a = z^a + \sum_{i=1}^N\left(u^a_i\hat{a}_i + u^a_i\hat{a}^{\dagger}_i\right),
\end{equation}
with $z^a=0$ for fermions. We will move back and forth between the $\hat{x},\hat{p}$ and $\hat{a},\hat{a}^\dagger$ bases via
\begin{equation}
    \hat{a} = \sqrt{\frac{\omega}{2}}\left(\hat{x} + \frac{i}{\omega}\hat{p}\right),
\end{equation}
in the following discussions\textemdash results are, of course basis independent.

\subsection{Example construction for a fermionic Gaussian state}\label{sec:Examples}
As a simple example, consider the ground state, $\ket{0}$, of the fermionic harmonic oscillator $\hat{H} = i\omega\hat{x}\hat{p} = \omega(\hat{a}^{\dagger}\hat{a} - \hat{a}\hat{a}^{\dagger})/2$. For the ease of calculation we will be working in the Fock basis so that the operator valued vector is $\hat{\xi}^a = (\hat{a},\hat{a}^{\dagger})$. The Hamiltonian can be written using the operator valued vector $\hat{\xi}^a$ in the following manner, $\hat{H} = \frac{1}{2}h_{ab}\hat{\xi}^a\hat{\xi}^b$. It follows that the Hamiltonian matrix $h_{ab}$ and the metric $G^{ab}$ for the system take the form,
\begin{equation}\label{eqn:FermionicOscGS_HandG}
    h_{ab} =
    \begin{pmatrix}
    0 & i\omega\\
    -i\omega &0
    \end{pmatrix}
    \text{ and }\hspace{0.05cm}
    G^{ab} = 
    \begin{pmatrix}
    0 & 1\\
    1 & 0
    \end{pmatrix},
\end{equation}
respectively. 
The state dependent symplectic form associated with the ground state of the fermionic harmonic oscillator is 
\begin{equation}\label{eqn:FermionicOscGS_Omega}
    \Omega^{ab} = 
    \begin{pmatrix}
    0 & -i\\
    i & 0
    \end{pmatrix}.
\end{equation}
And thus the linear complex structure is,
\begin{equation}\label{eqn:FermionicOscGS_J}
    J^a_b = \Omega^{ac}G_{cb}^{-1} = 
    \begin{pmatrix}
    -i & 0\\
    0 & i
    \end{pmatrix}.
\end{equation}
It is easy to see that $J^2 = -\mathds{I}$ implying that $G^{ab}$ and $\Omega^{ab}$ are K{\"a}hler compatible and thus the ground state of the fermionic harmonic oscillator is Gaussian. Given the form of the complex structure $J^a_b$ we can see that the gaussianity condition from (\ref{eqn:gausscondition}) takes on the form
\begin{equation}\label{eqn:FermionicOscGS_Annihil}
    \frac{1}{2}(\delta^a_b + iJ^a_b)\hat{\xi}^b\ket{0} = 
    \begin{pmatrix}
    \hat{a}\\
    0
    \end{pmatrix}\ket{0} = 0.
\end{equation}
%Since (\ref{eqn:FermionicOscGS_Annihil}) annihilates the ground state of the fermionic harmonic oscillator, the ground state is a Gaussian state. 
By similar means, we can also show that the first excited state of the fermionic harmonic oscillator, $\ket{1}$, with associated linear complex structure,
\begin{equation}
    J^a_b = 
    \left(\begin{array}{cc}
        i & 0 \\
        0 & -i
    \end{array}\right),
\end{equation}
is also a Gaussian state.

A general one-particle Gaussian state\footnote{We note that the most general Gaussian state can be constructed via action of the squeezing and displacement operators on a thermal state.} can be constructed by acting the squeezing operator
\begin{equation}\label{eqn:SqueezeOp}
    \hat{S}(r) = \exp\left[r\left(\hat{a}\hat{a} - \hat{a}^{\dagger}\hat{a}^{\dagger}\right)\right]
\end{equation}
and the displacement operator
\begin{equation}\label{eqn:DispOp}
    \hat{D}(\alpha) = \exp\left[\hat{a}^{\dagger}\gamma - \gamma^*\hat{a}\right].
\end{equation}
on either the ground state $\ket{0}$ or the first excited state $\ket{1}$. The squeezing operator $\hat{S}$ is parameterized by a squeezing parameter $r\in\mathds{R}$, where $r\to\infty$ indicates a highly localized state\footnote{We note that generally, the squeezing operator is a function of a complex parameter $\zeta = re^{i\phi}$, where $\phi$ is an arbitrary phase. For simplicity we have set $\phi = 0$.}.
Due to the Grassmann properties of fermionic states, the squeezing operator preserves both the ground and first excited state, \emph{i.e.} $\hat{S}(r)\ket{0} = \ket{0}$ and $\hat{S}(r)\ket{1} = \ket{1}$, respectively, for all values of the squeezing parameters $r$.
The squeezing operator is a unitary operator satisfying $\hat{S}(r)\hat{S}^{\dagger}(r) = \hat{S}^{\dagger}(r)\hat{S}(r) = \hat{\mathbf{I}}$.
The displacement operator $\hat{D}$ is parameterized by the variable $\gamma$, the amount of phase space displacement. For bosons $\gamma$ is a complex number and for fermions $\gamma$ is Grassmannian.
The action of the displacement operator on, specifically, the fermionic ground state or the first excited state produces a coherent state\footnote{A coherent state can be constructed by acting the displacement operator on the bosonic ground state as well.}, \emph{i.e.} $\hat{D}(\gamma)\ket{0} = \ket{\gamma}$ or $\hat{D}(\gamma)\ket{1} = \ket{\gamma}'$ for all values of the displacement parameter $\gamma$~\cite{CahillGlauber1999}. Like the squeezing operator, the displacement operator is also unitary, satisfying $\hat{D}(\gamma)\hat{D}^{\dagger}(\gamma) = \hat{D}^{\dagger}(\gamma)\hat{D}(\gamma) = \hat{\mathbf{I}}$. Given a fermionic Gaussian state $\ket{J}$, the action of the squeezing and displacement operators on the state, \emph{i.e.} $\hat{S}(r)\hat{D}(\gamma)\ket{J} = \ket{J'}$, produces a new fermionic Gaussian state $\ket{J'}$. For the proof of this statement see~\cite{HacklBianchi2021}, particularly Section $IIC4$ and Proposition 7.

\subsection{Localized states as squeezed Gaussian states}\label{sec:Squeezing}
In the position basis of a rigged Hilbert space, basis kets of the position operator, or states $\ket{x_0}$ such that $\hat{x}\ket{x_0}=x_0\ket{x_0}$, can be represented as a $\delta$-function: $\braket{x|x_0}=\delta(x-x_0)$.  A $\delta$-function is a limit of a sequence of normalized, narrowing Gaussians.  Reducing the width of a Gaussian is, however, simply squeezing the state in position.  Hence completely localized $\delta$-function states are equivalently Gaussian squeezed states in the infinite squeezing limit.  As discussed previously, Gaussian states can be constructed via action of the squeezing operator (\ref{eqn:SqueezeOp}) and displacement operator (\ref{eqn:DispOp}) on the Fock vacuum, \emph{i.e.} $\ket{J} = \hat{D}(\gamma)\hat{S}(r)\ket{0}$.  The position basis equivalence of such states to Gaussian functions centered around some $x_0$ and the equivalency between the position eigenkets $\ket{x'}$ and $\ket{J}$ in the minimal position uncertainty, infinitely squeezed limit is given in detail in equations 18-47 of~\cite{Munguia-GonzalezRegoFreericks2021}. The necessary result, which we present here is,
\begin{equation}\label{eqn:GaussianPosition}
    \ket{\chi_{x'}} = \left(\frac{\omega}{\pi}\right)^{1/4}\exp\left[-ix'\hat{p}\right]\exp\left[-\frac{1}{2}\left(\hat{a}^{\dagger}\right)^2\right]\ket{0}.
\end{equation}
The state $\ket{\chi_{x'}}$ is the Gaussian state that, in the infinitely squeezed limit, becomes a position basis state.  Since the momentum basis of a rigged Hilbert space shares the same properties as its position basis counterpart, the basis kets of the momentum basis can be written as shifted, squeezed Gaussian states, as shown in equation 48 of~\cite{Munguia-GonzalezRegoFreericks2021}, which we present below.
\begin{equation}\label{eqn:GaussianMomentum}
    \ket{\rho_{p'}} = \left(\frac{1}{\pi\omega}\right)^{1/4}\exp\left[ip'\hat{x}\right]\exp\left[\frac{1}{2}\left(\hat{a}^{\dagger}\right)^2\right]\ket{0}.
\end{equation}

An important consequence of the Gaussian description of the position basis is the structure of the inner product between basis kets $\ket{\chi_x}$ and $\ket{\chi_{x'}}$. Given the Gaussian description of these basis kets in (\ref{eqn:GaussianPosition}), the inner product may be written as,
\begin{equation}
    \braket{\chi_{x}|\chi_{x'}} = \left(\frac{\omega}{\pi}\right)^{1/2}\bra{\psi}\exp\left[-i(x'-x)\hat{p}\right]\ket{\psi},
\end{equation}
where we have defined the state $\ket{\psi} = \exp\left[-\left(\hat{a}^{\dagger}\right)^2/2\right]\ket{0}$. We can simplify the inner product by inserting a resolution of identity,
\begin{equation}\label{eqn:InnerProduct}
    \braket{\chi_{x}|\chi_{x'}} = \left(\frac{\omega}{\pi}\right)^{1/2}\int dp\text{ } e^{-i(x'-x)p}\braket{\psi|p}\braket{p|\psi}.
\end{equation}
To determine the projection of $\ket{\psi}$ onto the momentum basis states, it is easiest to write the momentum eigenkets as functions of the annihilation/creation operators acting on the Fock vacuum. The resulting state has the same form as (\ref{eqn:GaussianMomentum}), \emph{i.e.}, we can make the switch $\ket{p}\to\ket{\rho_p}$. It is a straightforward calculation to show that $\braket{\rho_p|\psi} =  (1/\pi\omega)^{1/2}\exp[-p^2/4\omega]$. Substituting this into (\ref{eqn:InnerProduct})  the inner product has the expected structure,
\begin{equation}\label{eqn:GaussianInnerProduct}
    \braket{\chi_x|\chi_{x'}} = \frac{1}{b\sqrt{\pi}}e^{-(x-x')^2/(2b)^2},
\end{equation}
where $b=(1/2\omega)^{1/2}$ is the width of the distribution. The inner product is a Gaussian in the difference of positions. In the limit where the width of the distribution becomes infinitesimally small, \emph{i.e.}, $b\to 0$, we recover the completely localized structure of the position basis kets.

%\\
%----------------------\\

\section{Fundamental frameworks}\label{sec:Fundamentals}
\subsection{Entanglement generation in a relational basis}\label{sec:RelationalBasis}
Given that $\ket{x}$ is the limit of a highly squeezed Gaussian state, we can use approximate position basis states without leaving the Gaussian framework. More importantly we can investigate how switching to a center-of-mass/relational partition affects entanglement of Gaussian states. Let us first consider two particles 1 and 2, localized at some points $x_1$ and $x_2$ respectively, with respect to some external reference frame.   We can write this state in an external partition as $\ket{\psi}=\ket{x_1} \otimes \ket{x_2}$.~\footnote{For the rest of this article, we will use $\ket{x}$ as a shorthand to refer to a highly localized Gaussian state around $x$, rather than an exact position eigenstate.} It is clear this state is a product state between the two particles without any entanglement.

In diffeomorphism invariant theories there is, of course, no external frame dependence.  One common approach to preserving diffeomorphism invariance in quantum mechanics is to move to a relational framework, where outcomes are defined in relation to others and probabilities become conditional (for a review, see~\cite{HohnSmithLock2021}).  In our framework we can construct a simple toy model that captures the relational aspect of a fully diffeomorphism invariant system by requiring that our quantum mechanical system be translationally invariant.  Intuitively, since we are dealing with position states, forcing a system to be translationally invariant will erase any absolute position information, leaving only translationally invariant relational degrees of freedom, such as $x_1-x_2$, in the reduced Hilbert space.  The first step on this path is to define new operators and a corresponding basis that capture the center-of-mass and relational degrees of freedom.  We will refer to this basis as the relational basis, in contrast to the external or absolute basis defined by the position states $\ket{x}$.

The states and operators with respect to the relational basis can be constructed via transformation from the external frame and partition. In the following, we follow the presentation in~\cite{Smith2017}. For a system with $N$ degrees of freedom, the position and momentum operators of the external partition, $\{\hat{x}_k,\hat{p}_k\}_{k=1}^N$ are fully specified by the center-of-mass position and momentum operators $\{\hat{x}_{cm},\hat{p}_{cm}\}$ and the relational position and momentum operators $\{\hat{x}_{i|1},\hat{p}_{i|1}\}_{i=2}^N$. The transformation between the positions and momenta of the external partition and the positions and momenta in the center-of-mass/relational partition are given by,
\begin{subequations}
\begin{align}
    \hat{x}_{cm} &= \frac{1}{M}\sum_{k=1}^n m_k\hat{x}_k,\\
    \hat{p}_{cm} &= \sum_{k=1}^n\hat{p}_k,\\
    \hat{x}_{i|1} &= \hat{x}_i - \hat{x}_1 \text{ for } i\in\{2,n\},\\
    \hat{p}_{i|1} &= \hat{p}_i - \tilde{m}_i\hat{p}_{cm} \text{ for } i\in\{2,n\},
\end{align}
\end{subequations}
where $M = \sum_{k=1}^n m_k$ is the total mass of the system and $\tilde{m}_i = m_i/M$ is the mass fraction of the $i$th particle. The canonical commutation relations for the operators in the center-of-mass/relational partition follow from the commutation relations between operators of the external partition, \emph{i.e.} $[\hat{x}_{cm},\hat{p}_{cm}] = [\hat{x}_{i|1},\hat{p}_{i|1}] = i$ with all other combinations vanishing. Notice that the relational position and momentum operators are defined with respect to particle 1, however any other particle may be chosen to the same effect.  

Transformations between the external partition and the center-of-mass/relational partition may engender entanglement if at least one of the particles is in a superposition in the external partition. As an example, consider a two particle composite state where, for simplicity, only particle one is in superposition with respect to the external partition\footnote{Note that we have chosen Gaussian states in the highly squeezed limit to allow for a trivial normalization. Generally, the Gaussian nature of the states does not allow for such a non-trivial normalizations since $\braket{x|x'}\neq\delta(x-x')$.},
\begin{equation}\label{eqn:TwoParticleExt}
\begin{aligned}
    \ket{\psi_{12}} &= \frac{1}{\sqrt{2}}\left(\ket{x_1} + \ket{x'_1}\right)\otimes\ket{x_2}\\
    &= \frac{1}{\sqrt{2}}\left(\ket{x_1}\otimes\ket{x_2} + \ket{x'_1}\otimes\ket{x_2}\right).
\end{aligned}
\end{equation}
Here $\ket{x_1},\ket{x'_1}\in\hil_1$ are the possible position states of particle 1 and $\ket{x_2}\in\hil_2$ is the position state of particle 2. It is clear no entanglement exists between degrees of freedom in the external partition. We now transform into the center-of-mass/relational partition, with particle 1 considered the ``reference'' particle from which the position of particle 2 will be defined.  Now the center-of-mass and relational position states for the two components of the superposition differ and $\ket{\psi}$ becomes
\begin{equation}\label{eqn:TwoParticleCM}
    \ket{\psi_{12}}\to\ket{\psi_{12,cm}} = \frac{1}{\sqrt{2}}\left(\ket{x_{cm}}\otimes\ket{x_{2|1}} + \ket{x'_{cm}}\otimes\ket{x'_{2|1}}\right),
\end{equation}
where the prime denotes the center-of-mass and relational position states between $\ket{x'_1}$ and $\ket{x_2}$. Now there is a bipartite entanglement between the center-of-mass and the relational position states~\cite{AngeloBrunnerPopescuShortSkrzypczyk2011}.

There is a subtle difference in the $n$-particle case when compared to the two-particle case, which we demonstrate with the three-particle case. Consider a three-particle composite state in the external frame where, for simplicity, only the reference particle is placed in superposition,
\begin{eqnarray}\label{eqn:ThreeParticleExt}\nonumber
    \ket{\psi_{123}} &=& \frac{1}{\sqrt{2}}\left(\ket{x_1} + \ket{x'_1}\right)\otimes\ket{x_2}\otimes\ket{x_3}\\\nonumber
    &=& \frac{1}{\sqrt{2}}\left(\ket{x_1}\otimes\ket{x_2}\otimes\ket{x_3} + \ket{x'_1}\otimes\ket{x_2}\otimes\ket{x_3}\right).\\
\end{eqnarray}
As in the previous example $\ket{x_1},\ket{x'_1}\in\hil_1$ are the possible position states of particle 1, the reference particle, $\ket{x_2}\in\hil_2$ is the position state of particle 2, and $\ket{x_3}\in\hil_3$ is the position state of particle 3. Currently there is no entanglement between any of the degrees of freedom. Different combinations of the states will generally have different centers-of-mass positions and different relational positions, however we would like to note that there exist cases where the center-of-mass and/or relational positions between two or more combinations are equal. After the transformation into the center-of-mass/relational partition, the state $\ket{\psi_{123}}$ is described as,
\begin{widetext}
\begin{equation}\label{eqn:ThreeParticleCM}
    \ket{\psi_{123}}\to\ket{\psi_{123,cm}} = \frac{1}{\sqrt{2}}\left(\ket{x_{cm}}\otimes\ket{x_{2|1}}\otimes\ket{x_{3|1}} + \ket{x'_{cm}}\otimes\ket{x'_{2|1}}\otimes\ket{x'_{3|1}}\right),
\end{equation}
\end{widetext}
where the primes denote center-of-mass and relational position states that utilized $\ket{x'_1}$ in their definitions.  Here there is tripartite entanglement between both the center-of-mass and relational degrees of freedom, which differs from the two particle case. In general, with different transformations and $n$-particles entanglement can be generated between various parts of the partitioned Hilbert space\textemdash one is not restricted to center-of-mass/relational bipartite entanglement.

Below we will extensively use the generalization of (\ref{eqn:TwoParticleExt}) and (\ref{eqn:ThreeParticleExt}) for calculational simplicity. For a system of $N$ particles, where one particle is placed in a location superposition with respect to the external frame, the state is given by 
\begin{widetext}
\begin{eqnarray}\nonumber
    \ket{\psi_{1,\hdots,N}} &=& \frac{1}{\sqrt{2}}\left(\ket{x_1} + \ket{x'_1}\right)\ket{x_2}\otimes\hdots\otimes\ket{x_N}\\
    &=& \frac{1}{\sqrt{2}}\left(\ket{x_1}\otimes\ket{x_2}\otimes\hdots\otimes\ket{x_N} + \ket{x'_1}\otimes\ket{x_2}\otimes\hdots\otimes\ket{x_N}\right)\label{eqn:NParticleExt}
\end{eqnarray}
\end{widetext}
where $\ket{x_1},\ket{x'_1}\in\hil_1$ are the possible position states of particle 1 and $\ket{x_i}\in\hil_i$ is the position state of the $i$th particle. The state in (\ref{eqn:NParticleExt}) written in the center-of-mass/relational partition is given by
\begin{equation}
\begin{aligned}
    \ket{\psi_{1,\hdots,N}}\to&\ket{\psi_{1,\hdots,N,cm}}\\ 
    &= \frac{1}{\sqrt{2}}\left(\ket{x_{cm}}\otimes\ket{x_{rel}} + \ket{x'_{cm}}\otimes\ket{x'_{rel}}\right)\label{eqn:NParticleCM}
\end{aligned}
\end{equation}
where $\ket{x_{rel}} = \ket{x_{2|1}}\otimes\hdots\otimes\ket{x_{N|1}}$ represents the $N-1$ remaining relational degrees of freedom. Notice that the state in (\ref{eqn:NParticleCM}) is an $n$-partite entangled state and our discussion regarding the nature of entanglement within the two partitions still holds.

\subsection{Removal of the external partition via a G-twirl over translations}\label{sec:Gtwirl}
The entangled states in the center-of-mass/relational partition are not yet fully relational, in that the center-of-mass coordinate contains degrees of freedom relative to the external partition.  To remove the center-of-mass degree of freedom we can group average over translations, which will reduce the state to one containing only relational degrees of freedom.  The procedure from quantum information theory for group averaging over quantum reference frames is known as the $G$-twirl~\cite{BartlettRudolphSpekkens2007}.

Consider a quantum state represented by the density matrix $\hat{\rho}\in\hil$ in some Hilbert space $\hil$, described with respect to some external reference frame. Changes to the orientation of the quantum state $\hat{\rho}$ with respect to the external reference frame are performed via the action of some unitary operation $\hat{U}(g)\in\hil$ on the state $\hat{\rho}$. Here, $\hat{U}(g)$ is the unitary representation a group element $g\in G$, where $G$ is the group of all possible changes of the external reference frame.
It is important to note that in the definition of the $\mathcal{G}$-twirl, $G$ is a compact group~\cite{BartlettRudolphSpekkens2007}, however as we will see $G$ can be non-compact albeit yielding slightly more complex results~\cite{SmithPianiMann2016}. The result of the $G$-twirl is the description of the quantum state that does not contain any information about the external frame. This description is achieved by averaging over all possible orientations of $\hat{\rho}$ with respect to the external frame, where every possible orientation is weighted equally,
\begin{equation}
    \hat{\rho}_R = \mathcal{G}[\hat{\rho}]\equiv\int dg\hat{U}(g)\hat{\rho}\hat{U}^{\dagger}(g),
\end{equation}
where $dg$ is the Haar measure of the group $G$ and $\hat{\rho}_R\in\hil$ is the relational description of $\hat{\rho}$.

Since we are averaging over all elements of the group that transforms the external frame, we are removing any relation to the external reference frame that was used to describe the state $\hat{\rho}$. Only the relational degrees of freedom within the system remain, \emph{i.e.} the information unaffected by changes in the external reference frame. For example, suppose $\hat{\rho}$ describes a composite state of two particles such that $\hil = \hil_1\otimes\hil_2$. After a $G$-twirl is performed, the remaining information contains the relational degrees of freedom between the two particles. Note that the $G$-twirl is done via the product representation $\hat{U}(g) = \hat{U}_1(g)\otimes\hat{U}_2(g)$, where $\hat{U}_1(g)\in\hil_1$ and $\hat{U}_2(g)\in\hil_2$ are the unitary representations of the group $G$ in each Hilbert space.

Since we are interested in translation invariant states, we will focus on the $G$-twirl operation as it relates to translations. Our presentation below follows from the work of~\cite{Smith2017}. For more details we encourage the reader to review the original work. The action of the three-dimensional translation group $g = x\in\mathbb{R}$ on the external frame in the external partition, $\hil = \bigotimes^N_{n=1}\hil_n$ is given by
\begin{equation}
    \hat{U}(x) = \bigotimes^N_{n=1}e^{-ix\hat{p}_n}.
\end{equation}
In the center-of-mass/relational partition, $\hil_{cm}\otimes\hil_R$, the action of the translation group is given by
\begin{equation}
    \hat{U}(x) = e^{-ix\hat{p}_{cm}}\otimes\hat{\mathbf{I}}_R.
\end{equation}

To carry out the $G$-twirl over translations we first express the state $\hat{\rho}$ in the center-of-mass/relational partition, in the momentum basis,
\begin{widetext}
\begin{equation}
    \hat{\rho} = \int dp_{cm}dp'_{cm}dp_{R}dp'_{R}\text{ }\rho(p_{cm},p'_{cm},p_R,p'_R)\ket{p_{cm}}\bra{p'_{cm}}\otimes\ket{p_R}\bra{p'_R},
\end{equation}
where $\ket{p_{cm}}$ and $\ket{p'_{cm}}$ are the possible basis states of the center-of-mass momentum and similarly, $\ket{p_R}$ and $\ket{p'_R}$ are the basis states of the $N-1$ relational momentum vectors. It follows that the $G$-twirl over the possible translations of the external frame is given by
\begin{eqnarray}\nonumber
    \mathcal{G}_T[\hat{\rho}] &=& \int dx\hat{U}(x)\left[\int dp_{cm}dp'_{cm}dp_{R}dp'_{R}\rho(p_{cm},p'_{cm},p_R,p'_R)\ket{p_{cm}}\bra{p'_{cm}}\otimes\ket{p_R}\bra{p'_R}\right]\hat{U}^{\dagger}(x)\\\nonumber
    &=& \int dxdp_{cm}dp'_{cm}dp_{R}dp'_{R}\rho(p_{cm},p'_{cm},p_R,p'_R) e^{-ix\hat{p}_{cm}}\ket{p_{cm}}\bra{p'_{cm}}e^{ix\hat{p}'_{cm}}\otimes\ket{p_R}\bra{p'_R}\\
    &=& (2\pi)^3 \int dp_{cm}dp_{R}dp'_{R}\text{ }\rho(p_{cm},p_{cm},p_R,p'_R) \ket{p_{cm}}\bra{p_{cm}}\otimes\ket{p_R}\bra{p'_R}\label{eqn:PositionGtwirl}.
\end{eqnarray}
\end{widetext}
Going from the second to the third equality we have used the definition of the Dirac delta function $(2\pi)^3\delta(p - p') \equiv \int dxe^{ix(p-p')}$. It is clear that the $G$-twirl operation is effectively the trace over the center-of-mass degrees of freedom in the center-of-mass/relational partition, \emph{i.e.} $\mathcal{G}_T[\hat{\rho}] = \mathcal{I}\Tr_{cm}\hat{\rho}$, where $\mathcal{I}$ denotes a divergent constant originating from integral over the probability function $\rho(p_{cm},p_{cm},p_R,p'_R)$ in (\ref{eqn:PositionGtwirl}). The new state $\mathcal{G}_T[\hat{\rho}]$ is not normalized, since there are infinitely many states to trace over, due in part to the noncompact nature of the group of translations. However, the divergent nature of $\mathcal{I}$ will be inconsequential to the physics of interest for reasons we will discuss below.

As an example, let us perform a $G$-twirl on the state in (\ref{eqn:TwoParticleCM}) over the set of one-dimensional translations. Since the relational state has been shown to be $\mathcal{G}_T[\hat{\rho}] = \mathcal{I}\Tr_{cm}\hat{\rho}$ it follows that
\begin{widetext}
\begin{equation}
    \mathcal{G}_T[\ket{\psi_{12,cm}}\bra{\psi_{12,cm}}] = \mathcal{I}\Tr_{cm}[\ket{\psi_{12,cm}}\bra{\psi_{12,cm}}] = \mathcal{I}\left(\ket{x_{2|1}}\bra{x_{2|1}} + \ket{x'_{2|1}}\bra{x'_{2|1}}\right).
\end{equation}
Here the primes refer to center-of-mass and relational states defined with $\ket{x'_1}$ in (\ref{eqn:TwoParticleExt}). We note that any normalization factors are absorbed into $\mathcal{I}$. Now we have a completely relational state, one that only contains relational information and no information regarding the external frame. Furthermore, the entanglement between the center-of-mass and the relational degrees of freedom has been removed and the relational degrees of freedom are now in a mixed state. Notice that the $G$-twirl destroys the entanglement that existed between the center-of-mass and relational degrees of freedom.

A slightly more interesting example is the $G$-twirl of the three particle state in (\ref{eqn:ThreeParticleCM}), over the set of one-dimensional translations. It follows that,
\begin{eqnarray}\nonumber
    \mathcal{G}_T[\ket{\psi_{123,cm}}\bra{\psi_{123,cm}}] &=& \mathcal{I}\Tr_{cm}[\ket{\psi_{123,cm}}\bra{\psi_{123,cm}}]\\
    &=& \mathcal{I}\left(\ket{x_{2|1}}\bra{x_{2|1}}\otimes\ket{x_{3|1}}\bra{x_{3|1}} + \ket{x'_{2|1}}\bra{x'_{2|1}}\otimes\ket{x'_{3|1}}\bra{x'_{3|1}}\right).
\end{eqnarray}
\end{widetext}

Even though the original pure state was tripartite entangled, the $G$-twirl has removed all the entanglement from the system. This is clear since the post $G$-twirl state is a separable, mixed state over the center-of-mass/relational partition. Moreover, this implies that constructing our purely relational state by discarding information about the external frame removes any entanglement as well. As we will demonstrate below, this phenomena makes for an interesting, case study of the entanglement extraction protocol as applied to relational degrees of freedom.

\subsection{``Relationalizing'' of the external partition via the $Z$-model}\label{sec:Zmodel}
The labeling of spacetime points provided by an external frame has no physical meaning. This implies that states in the external partition don't come from any specific measurement of an explicitly defined, physical observable, rather they can be thought of as a choice of a particular gauge. As we have shown above, the $G$-twirl effectively removes this gauge choice, by integrating out a state's dependence on the external frame, leaving only relational information behind.

Alternatively, one can put the external partition into a relational framework by correlating each position state $\ket{x}$ with some dynamical observable. This will give the arbitrary labeling meaning since the position states are now linked to the value of a dynamical observable. However, we must choose such observables carefully. For example, suppose we chose the Hamiltonian $\hat{H}$ to be our "reference" observable. Since $\hat{H}$ is fundamentally a generator of time translations, it has no intrinsic dependence on position. This allows us to choose $\hat{H}$ to be translationally invariant, which puts us back into the $G$-twirl scenario described in the previous section. To give $\hat{H}$ position dependence we will employ the $Z$-model construction. Below we will present a brief overview of the construction and for more details we encourage the reader to view the original work~\cite{GiddingsMarolfHartle2006}.

Given a particular quantum state, the $Z$-model allows us to define the location of a local observable by specifying it relative to a structure determined by the expectation value of a pseudo-local observable. This allows us to give physical meaning to the labeling provided by the external partition. Since we are in one spatial dimension, we can introduce a dynamical, massless, auxiliary field $\hat{Z}$, and a state $\ket{\psi_Z}$. We then define
\begin{equation}
 \lambda x= \bra{\psi_Z}\hat{Z}\ket{\psi_Z},
\end{equation}
where $\lambda$ is a proportionality constant. Here, we have defined our spatial coordinate $x$ in terms of the expectation values of the auxiliary field, which obey the classical equations of motion. Since any coordinate system is monotonically increasing, the expectation value of the auxiliary field must also be monotonically increasing, implying that the gradient of the auxiliary field over the spacetime is everywhere nonvanishing.

We can add position dependence into the Hamiltonian by including a $Z$-model coupling term, \emph{e.g.} in the context of field theory a $\hat{H}_{int} = \psi(x)Z(x)$ term, where $\psi(x)$ is a field of the original model, and $Z(x)$ is the auxiliary field used to define the external frame. Since the auxiliary field is dynamical and spontaneously breaks the translation invariance of the Hamiltonian, the fundamental underlying translational invariance of the theory remains intact.  
%Only when the $Z$ field has an everywhere non-vanishing gradient, which is the property of a solution, can we establish an actual spatial coordinate system.

To concretely implement this idea within our framework, consider a system within a parallel plate capacitor. A system of charged particles, with individual charge $q$ located within the capacitor, will be represented by a family of Gaussian states. The plates, separated by distance $L$ and with charge density $\pm \sigma$ on the left/right plates respectively, will produce an everywhere (inside the plates) non-vanishing electric field $E$. The potential field therefore has a non-vanishing gradient everywhere within the capacitor. Therefore the $Z$-model can be incorporated into our framework via the usual electromagnetic coupling, which does not affect the Gaussian nature of the allowed states of the theory due to the form on the interaction term. Since the potential field has a non-vanishing gradient between the plates, it will be monotonic within the capacitor, allowing us to map the position $x$ to the value of the electric potential of a particle at point $x$, \emph{i.e.} the expectation value of the interaction term in the Hamiltonian. Particularly, the energy of a Gaussian state, centered around $x$ is given by,
\begin{equation}\label{eqn:EnergyPositionRelation}
    \braket{\hat{H}_{int}(x)} = q\sigma\braket{x},
\end{equation}
the expectation value of the electromagnetic coupling term.

Consider the three particle state in the center-of-mass/relational partition shown in (\ref{eqn:ThreeParticleCM}). When determining the energy expectation value of this state in the $Z$-model, we find that, while the relational position eigenstate is independent of the electric potential, the center-of-mass eigenstate is not. Hence, there is a 1-to-1 correspondence between $x_{cm}$ and the observable $\hat{H}_{int}(x_{cm})$. Therefore, we are unable to remove the information about the external frame. Moreover, since the coupling is with the expectation value then as long as we assume there is never entanglement between the auxiliary field and the Gaussian states we can think about entanglement extraction between the Gaussian states themselves. In other words, as long as the auxiliary field is coupled semi-classically, the entanglement extraction analysis for relational systems is still possible without worrying about $Z$-field entanglement.
(However, we note that this assumption can be relaxed and investigated further in any future work.)

\subsection{Entanglement Extraction Protocol and Energy Cost}\label{sec:Extraction}
Fundamentally, the entanglement extraction process requires a source system $S$ that contains entangled states, and two target systems that are unentangled. While the location of these systems is generally unimportant, we have chosen to locate the target systems at spatial infinity. The source system is composed of two localized subsystems $A$ and $B$ with associated Hilbert spaces $\hil_A$ and $\hil_B$, respectively. Similarly, the two target systems $1$ and $2$ have associated Hilbert spaces $\hil_1$ and $\hil_2$, respectively. Since the source system is composed of local subsystem Hilbert spaces the composite Hilbert space is factorizable, \emph{i.e.} $\hil_S = \hil_A\otimes\hil_B$, and similarly for the composite Hilbert space of the target systems after the extraction of entanglement from the source system to the target systems has occurred, \emph{i.e.} $\hil_F = \hil_1\otimes\hil_2$. 
Factorizability allows one to define entanglement on the composite Hilbert spaces.

Suppose the source system contains a set of entangled states and the target systems each contain one state\footnote{Such states can be either bosonic or fermionic.}. Initially, the total system is in the product state,
\begin{equation}\label{eqn:InitialState}
    \hat{\rho}_I = \hat{\rho}_1\otimes\hat{\rho}_2\otimes\ket{\Psi}_S\bra{\Psi}_S,
\end{equation}
where $\ket{\Psi}_S$ is the initial, entangled state of the source system and $\hat{\rho}_1, \hat{\rho}_2$ are the initial density matrices of the two target systems. 
Entanglement is extracted by ``swapping'' the unentangled modes of the target system with the entangled modes of the source system.  Once the swapping procedure is complete the source subsystems $A$ and $B$ are unentangled and the target systems $1$ and $2$ are now entangled. 

Entanglement is extracted from the source system by a set of ``swap'' operations\textemdash two unitary operations $\hat{U}_1$ and $\hat{U}_2$ that map one target mode to one mode inside the source system~\cite{HacklJonsson2019}. For example, the target modes, $(\hat{a}_1, \hat{a}_1^{\dagger})$ and $(\hat{a}_2, \hat{a}_2^{\dagger})$, are swapped with one mode inside the source system, $(\hat{a}_A, \hat{a}_A^{\dagger})$ and $(\hat{a}_B, \hat{a}_B^{\dagger})$, using the following relations,
\begin{equation}\label{eqn:SwappingTransforms}
\begin{aligned}
    \hat{U}^{\dagger}_1\hat{a}_1\hat{U}_1 = \hat{a}_A, \hspace{1cm} &\hat{U}^{\dagger}_1\hat{a}_A\hat{U}_1 = \hat{a}_1,\\
    \hat{U}^{\dagger}_1\hat{a}_1^{\dagger}\hat{U}_1 = \hat{a}_A^{\dagger}, \hspace{1cm} &\hat{U}^{\dagger}_1\hat{a}_A^{\dagger}\hat{U}_1 = \hat{a}_1^{\dagger},
\end{aligned}
\end{equation}
while $\hat{U}_2$ swaps $(\hat{a}_2, \hat{a}_2^{\dagger})$ and $(\hat{a}_B, \hat{a}_B^{\dagger})$. Note that these types of unitary operations always exist due to bit symmetry~\cite{MullerUdudec2012}. After the swap operations, the two target systems become entangled. The entanglement content of the state depends on the type of source and target systems, their couplings, and resources such as energy available to implement the extraction. To ensure the entanglement of the two target modes was preexisting in the system and not created by the swap operations, the source modes are restricted so they either commute or anti-commute depending on whether the source system is a set of bosons or fermions, respectively. For example, suppose the source system is a set of entangled fermions. It follows that the restrictions on the fermionic modes are,
\begin{equation}\label{eqn:ExtractionCondition}
    [\hat{a}_A,\hat{a}_B]_+=[\hat{a}_A,\hat{a}_B^{\dagger}]_+=[\hat{a}_A^{\dagger},\hat{a}_B]_+=[\hat{a}_A^{\dagger},\hat{a}_B^{\dagger}]_+=0.
\end{equation}
The conditions like those in (\ref{eqn:ExtractionCondition}) ensure that the entanglement of the two target modes was not created by the swap operations between the target and source modes~\cite{SimidzijaJonssonMartinMartinez2018}.

An initial configuration can be chosen such that the source system is a product state between the two modes and the rest of the system
\begin{equation}\label{eqn:InitialSystemState}
    \ket{\Psi}_S = \ket{\psi}_{AB}\otimes\ket{\phi}_R,
\end{equation}
where $\ket{\psi}_{AB}$ is the initial entangled state comprised of states localized to Hilbert space $\hil_A$ and $\hil_B$, and $\ket{\phi}_R$ is the ground state for the rest of the system. 
When the target modes and source modes are swapped the total system is placed in the state
\begin{equation}\label{eqn:FinalState}
    \hat{\rho}_F = \ket{\psi}\bra{\psi}_{12}\otimes\hat{\rho}_A\otimes\hat{\rho}_{B}\otimes\ket{\phi}\bra{\phi}_R,
\end{equation}
where $\ket{\psi}_{12}$ is the final, entangled state of the two target systems and $\hat{\rho}_A$ and $\hat{\rho}_B$ are the final states of the two subsystems $A$ and $B$ that compose the source system. 
Notice that the rest of source system is unaffected by the extraction process and remains in the ground state $\ket{\phi}_R$.

Hackl and Jonsson assume that the initial source system $S$ possesses entanglement in its ground state.  Hence after any entanglement extraction the states $\hat{\rho}_A$ and $\hat{\rho}_B$ must be in a higher energy state.
It follows that the energy expectation value has increased due to the extraction process. 
This is the cost of the entanglement extraction.
For more details see~\cite{HacklJonsson2019}.

The energy cost of entanglement extraction is given by the difference between the expectation value of the source system's Hamiltonian before and after the extraction, \emph{i.e.}
\begin{equation}\label{eqn:GeneralEnergyCost}
    \Delta E = \Tr\left(\hat{\rho}_A\otimes\hat{\rho}_{B}\otimes\ket{\phi}\bra{\phi}_R\hat{H} - \ket{\Psi}\bra{\Psi}_S\hat{H}\right).
\end{equation}
The Hamiltonian $\hat{H}$ of the source system may be coupling different modes of the system, however for the calculation of the energy cost, only the parts acting on the two modes $A,B$ are relevant. Therefore, the energy cost of entanglement extraction is solely determined by $\hat{H}_{AB}$ acting on only the $A,B$ modes, \emph{i.e.}
\begin{equation}\label{eqn:PartnerEnergyCost}
    \Delta E = \Tr\left(\hat{\rho}_A\otimes\hat{\rho}_{B}\hat{H}_{AB}\right) - \Tr\left(\ket{\psi}\bra{\psi}_{AB}\hat{H}_{AB}\right)
\end{equation}

If one wishes to minimize the extraction energy cost then two requirements need to be fulfilled. First, one must choose the source modes to be partner modes, \emph{i.e.} $\hat{a}_B = \hat{a}_{\bar{A}}$ and $\hat{a}^{\dagger}_B = \hat{a}^{\dagger}_{\bar{A}}$. Such a choice maximizes the extracted entanglement. If non-partner modes are chosen, then the mixed state entanglement between the target modes is never lager than between the the mode and its partner. Second, the target modes need to be initialized in the ground states of the single-mode restrictions $\hat{H}_A$ and $\hat{H}_{B}$ of $\hat{H}$ onto the individual partner modes.
For bosonic and fermionic modes, these states are Gaussian. For more details see~\cite{HacklJonsson2019}.

For every mode in a given multi-particle Gaussian state system, there exists a mode that shares all of the first mode's entanglement~\cite{HacklJonsson2019}. This implies that one can always find partner modes within multi-particle Gaussian systems. Thus, as long as the entanglement extraction protocol can be performed, one can always choose partner modes to minimize the energy cost of entanglement extraction. If one does not choose partner modes, the energy cost of entanglement extraction is not necessarily minimized. In the work below, we do not explicitly choose partner modes when we perform the entanglement extraction protocol. Thus, our calculated energy cost is not necessarily minimized. Minimizing the energy cost is beyond the scope of this work since we merely seek to demonstrate when there is an energy cost to this processes for various relational approaches. Whether or not that cost is minimized is left for future work.
%\\
%----------------------\\

\section{Entanglement extraction in relational systems}\label{sec:RelationalSystems}
We now turn to the process of entanglement extraction for relational systems, using the pieces developed in the previous section. In particular we show how the entanglement extraction protocol from Section \ref{sec:Extraction} can be implemented alongside the $G$-twirl from section \ref{sec:Gtwirl} and the $Z$-model from Section \ref{sec:Zmodel}. We also demonstrate how to smoothly transition between these concepts.

\subsection{Lack of entanglement extraction in G-twirled relational partitions}\label{sec:Gextraction}
In Section \ref{sec:RelationalBasis} we demonstrated that a pure, non-entangled state in some external partition can become entangled by writing the state in the center-of-mass/relational partition~\cite{SmithPianiMann2016,Smith2017}. By writing the state in the center-of-mass/relational partition, we have entanglement between the center-of-mass degrees of freedom and the relational degrees of freedom. For examples see (\ref{eqn:TwoParticleCM}), (\ref{eqn:ThreeParticleCM}), or (\ref{eqn:NParticleCM}). However, the transformation into the center-of-mass/relational partition does not produce a purely relational state since the center-of-mass degrees of freedom, which couple to the external frame, still exist. For our purposes, we are interested in the energy cost of entanglement extraction for purely relational degrees of freedom. The absolute position of the center-of-mass degrees of freedom, in this case $x_{cm}$, is unknown without specifying some measurement system (reference frame) that is capable of differentiating different $x_{cm}$'s. However, if $x_{cm}$ is translationally invariant, then such a measurement system cannot exist. Therefore the center-of-mass degrees of freedom are not measurable degrees of freedom, rather they are gauge degrees of freedom. We can remove these gauge degrees of freedom by $G$-twirling over the center-of-mass degrees of freedom. As we showed in Section \ref{sec:Gtwirl} the $G$-twirl will produce a purely relational state, but at the cost of destroying the entanglement within the center-of-mass and relational degrees of freedom. After the $G$-twirl is done, the relational degrees of freedom are left in a mixed state that is not entangled. Therefore there is no possibility of an energy cost of entanglement extraction\textemdash the entanglement extraction protocol cannot even be performed.

\subsection{Entanglement extraction via the $Z$-model}\label{sec:Zextraction}
In Section \ref{sec:Zmodel} we demonstrated that the $Z$-model tied the absolute position of a state with the configuration of an auxiliary field. This implies that the coordinate system used to define the absolute position is no longer a gauge degree of freedom. Furthermore, since the coordinate system now depends on the configuration of auxiliary degree of freedom, the external partition has been ``relationalized''. We can now examine the entanglement extraction protocol from Section \ref{sec:Extraction} for our purely relational state. This is in contrast to the use of the $G$-twirl, where the $G$-twirl process creates a relational state by destroying the entanglement between the relational and non-relational parts of the state.

In our framework with the $Z$-model, the absolute position of charged particles is given meaning by the gradient electromagnetic vector potential inside a parallel plate capacitor. This means that the two location configurations in the superposition found in (\ref{eqn:TwoParticleExt}), (\ref{eqn:ThreeParticleExt}), or (\ref{eqn:NParticleExt}) will have different energies. As a simple example, consider the superposition state found in (\ref{eqn:TwoParticleExt}). Since $\ket{x_1}$ and $\ket{x'_1}$ will interact differently with the $Z$-model coupling in the Hamiltonian, the energy for the components of the superposition state would be,
\begin{eqnarray}
    E &=& \bra{x_2}\otimes\bra{x_1}\hat{H}_{12}\ket{x_1}\otimes\ket{x_2},\\
    E' &=& \bra{x_2}\otimes\bra{x'_1}\hat{H}_{12}\ket{x'_1}\otimes\ket{x_2}.
\end{eqnarray}
Since $\ket{x_1}$ and $\ket{x'_1}$ are now physically different states, $E\neq E'$. The same principle is applicable to $n$ particles, where at least one particle in a location superposition. Furthermore, these energy differences are unchanged by the choice of partition of the Hilbert space. Therefore, we will see the same effect in the center-of-mass/relational partition as well. Since there is entanglement and an energy difference we are able to perform the entanglement extraction protocol as prescribed and find a non-vanishing energy cost. 

Let the $n$-particle initial state in the center-of-mass/relational partition from (\ref{eqn:NParticleCM}) be the initial entangled state of the source system,
\begin{equation}
    \ket{\psi_I} = \ket{x_{cm}}\otimes\ket{x_{rel}} + \ket{x'_{cm}}\otimes\ket{x'_{rel}}
\end{equation}
is relational. This state written as a density matrix is $\hat{\rho}_I = \ket{\psi_I}\bra{\psi_I}$. After the extraction process, the system will be left in a mixed state
\begin{widetext}
\begin{eqnarray}
    \hat{\rho}_F = \left(\ket{x_{cm}}\bra{x_{cm}} + \ket{x'_{cm}}\bra{x'_{cm}}\right)\otimes\left(\ket{x_{rel}}\bra{x_{rel}} + \ket{x'_{rel}}\bra{x'_{rel}}\right) = \hat{\rho}_{cm}\otimes\hat{\rho}_{rel}
\end{eqnarray}
\end{widetext}
In Hackl and Jonson's work the final state of the source system is a statistical mixture of states. As such, the final energy will be a statistical average of the energy for the state $\ket{x_{cm}}\otimes\ket{x_{rel}}$ and the state $\ket{x'_{cm}}\otimes\ket{x'_{rel}}$. Since we are only considering one state, our final state will not be mixed but rather be either the primed or the unprimed state. 

Unlike the scenario of Hackl and Jonsson, the initial state is not in the ground state by definition.  In order to even run the $Z$-model, we needed to correlate the position with different energies via the Hamiltonian interaction.  This means that any state, other than the single ``ground state'' position state at the location of one of the plates, has a higher energy.  Hence the extraction process won't yield a minimum energy, but instead the energy difference, as that was the resource we used in the $Z$-model to establish the non-gauge nature of the external partition. The energy difference is given by
\begin{equation}
    \Delta E = \Tr\left[\hat{\rho}_{cm}\otimes\hat{\rho}_{rel}\hat{H}\right] - \Tr\left[\hat{\rho}_I\hat{H}\right].
\end{equation}

No matter the final state, the energy difference is non-vanishing and so there will be a net energy transfer to/from the target modes for extracting entanglement from this relational system.  Since the final state will be measured in either the primed or unprimed location, there is hence an energy change for localizing the relational quantum information contained in the state.  Importantly, we note the connection between entanglement extraction and how we made the system relational.  In the first approach, when we throw out any local information, there is no way to perform any quantum information process related to localization or entanglement extraction at the end.  And indeed, we saw that the protocol fails.  In the second, when the location is made relational via the $Z$-model, the necessary change to the Hamiltonian also automatically enables the extraction protocol to occur.  The two processes were locked together.

Of course, these results are only possible inside the capacitor. Outside the capacitor, where the electric field used for the $Z$-model does not exist the entanglement extraction protocol does not work, since there is no way to ``relationalize'' the states. This implies that there is a limited domain where the entanglement extraction protocol works. This is appropriate for a relational set-up with finite experimental configurations. Outside the capacitor it is required to use the $G$-twirl to construct relational states, since there is no auxiliary with a monotonic gradient to use for the $Z$-model. The question then becomes, how can we smoothly transition between the $Z$-model with the entanglement extraction protocol to the $G$-twirl without the entanglement extraction protocol? We show below that this smooth transition can be carried out via positive operator valued measurements (POVM).

\subsection{From $G$ to $Z$ via POVM}\label{sec:GtoZ}
From the time-energy uncertainty relation, we know that the time it takes to make a measurement scales as $\Delta t \sim 1/\Delta E$, where $\Delta E$ is the uncertainty in the energy of the system~\cite{DeffnerCampbell2017}. Thus, any physically realizable system has some innate uncertainty in its energy. Therefore, resolving the position in a finite time has some inherent inaccuracy since,
\begin{equation}\label{eqn:PositionUncertainty}
    \braket{\Delta x} \sim \frac{\braket{\Delta \hat{H}_{int}(x)}}{q\sigma}.
\end{equation}
The expression in (\ref{eqn:PositionUncertainty}) is a result of considering the uncertainty in position defined in the $Z$-model, \emph{i.e.} the uncertainty of (\ref{eqn:EnergyPositionRelation}). It is clear from the above relationship that when the charge on the plates of the capacitor decreases, the uncertainty in $x$ increases. In other words, there must be a smooth limit where the $Z$-model produces complete uncertainty about the position. If we were to keep only relational degrees of freedom, this limit should also reproduce the $G$-twirl. We can construct the transition via the framework of POVM.

As a reminder, POVM are a set of positive semi-definite Hermitian matrices $\{\hat{P}_m\}$, where $m$ is the value of each measurement, on a Hilbert space $\hil$, that sums to the identity, \emph{i.e.} for every measurement $m$,
\begin{equation}
    \sum_m \hat{P}_m = \hat{\mathbf{I}}.
\end{equation}
Generally the exact form of each $\hat{P}_m$ is unknown, however for our purposes it is convenient to assume the operators describe perfect measurements, \emph{i.e.} $\hat{P}_m = \ket{m}\bra{m}$, where $\ket{m}$ is a measurement eigenstate with eigenvalue $m$. Given a pure state, $\ket{\psi}$ and a set of POVM $\hat{P}_m$, the probability $\ket{\psi}$ is in the $\ket{m}$ state when measured is,
\begin{equation}
    p(m) = \bra{\psi}\hat{P}_m\ket{\psi} = \Tr(\hat{\rho}\hat{P}_m),
\end{equation}
where $\hat{\rho} = \ket{\psi}\bra{\psi}$ is the associated density matrix. For mixed states, the probability is solely given by the trace over the product of the density matrix and the POVM operators.

Through the lens of POVM, the energy expectation value of the detector, given a particular center-of-mass of the particle system is given by,
\begin{equation}
    \braket{\hat{H}_{int}(x)} = \sum_j\mathcal{E}_j(x_{cm})p(x_{cm})
\end{equation}
where $\mathcal{E}_j(x_{cm})$ is a map between the center-of-mass of the system and the energy read by the detector. We assume that the detector has a minimum uncertainty. That is, we assume that when the detector makes an energy measurement, the true energy of the system is placed into bins $\mathcal{E}_j(x_{cm})$. The true energy of the system is then within the bin energy and the energy uncertainty, \emph{i.e.} $\mathcal{E}_j(x_{cm}) + \Delta E$. The probability of measuring the state with a center-of-mass of $x_{cm}$ within one particular bin is given by,
\begin{equation}
    p(x_{cm}) = \int_{x_{cm,i}}^{x_{cm,i}+\Delta x} dx''_{cm} \Tr\left[\ket{x_{cm,i}}\braket{x_{cm,i}|\psi}\bra{\psi}\right].
\end{equation}
The notion of a perfect measurement is given by $\hat{P}_{x_{cm}} =\ket{x_{cm,i}}\bra{x_{cm,i}}$. The state $\ket{\psi}$ we will take to be the $n$-particle entangled state in the center-of-mass/relational partition from (\ref{eqn:NParticleCM}). Assuming the trace is done over the infinite set of center-of-mass position states, the probability can be written as
\begin{widetext}
\begin{eqnarray}\nonumber
        p(x_{cm}) &=& \int_{x_{cm,i}}^{x_{cm,i}+\Delta x_{cm}} dx''_{cm} \left[\braket{x''_{cm}|x_{cm}}\braket{x_{cm}|x''_{cm}}\ket{x_{rel}}\bra{x_{rel}} + \braket{x''_{cm}|x_{cm}}\braket{x'_{cm}|x''_{cm}}\ket{x_{rel}}\bra{x'_{rel}}\right.\\
    &&\hspace{3cm} + \braket{x''_{cm}|x'_{cm}}\braket{x_{cm}|x''_{cm}}\ket{x'_{rel}}\bra{x_{rel}} \left. + \braket{x''_{cm}|x'_{cm}}\braket{x'_{cm}|x''_{cm}}\ket{x'_{rel}}\bra{x'_{rel}}\right].\label{eqn:ProbMeasure}
\end{eqnarray}

In the limit where the charge on the plates goes to zero, $q\sigma\to 0$, the uncertainty of the position measurement becomes infinite, $\Delta x\to 0$. This implies that the integral in (\ref{eqn:ProbMeasure}) is now over all space instead of two arbitrary bins of the detector. Since the integral is over all space, it simplifies to
\begin{eqnarray}
        p(x_{cm}) &=& \braket{x_{cm}|x_{cm}}\ket{x_{rel}}\bra{x_{rel}} + \braket{x'_{cm}|x_{cm}}\ket{x_{rel}}\bra{x'_{rel}} + \braket{x_{cm}|x'_{cm}}\ket{x'_{rel}}\bra{x_{rel}} + \braket{x'_{cm}|x'_{cm}}\ket{x'_{rel}}\bra{x'_{rel}}.
\end{eqnarray}
The results of the inner products are Gaussians given by (\ref{eqn:GaussianInnerProduct}). If we consider the highly localized limit, \emph{i.e.} $b\to0$, the inner products become delta functions. It is easy to see that since $x_{cm}\neq x'_{cm}$ the second and third terms will vanish and the first and fourth terms will remain, albeit with a divergent coefficient. It follows that,
\begin{equation}
    p(x_{cm}) = \mathcal{N}\left(\ket{x_{rel}}\bra{x_{rel}} + \ket{x'_{rel}}\bra{x'_{rel}}\right),
\end{equation}
where $\mathcal{N}$ is divergent. Notice that this result matches the $G$-twirl result from Section \ref{sec:Gtwirl}.

In the limit where the charge on the plates becomes strong, $q\sigma\to\infty$, the uncertainty of the position measurement becomes zero $\Delta x\to 0$. However, for our purposes is is sufficient to consider $\Delta x$ to be finite. As a consequence we cannot utilize the resolution of identity as we did previously since the integral is over a finite subset of $x_{cm}$. Using the inner product from (\ref{eqn:GaussianInnerProduct}) the integral becomes,
\begin{eqnarray}\nonumber
    p(x_{cm}) &=& \frac{1}{b^2\pi}\int_{x_{cm,i}}^{x_{cm,i}+\Delta x_{cm}} dx''_{cm} \left[e^{-(x_{cm}''-x_{cm})^2/2b^2}\ket{x_{rel}}\bra{x_{rel}}\right. + e^{-(x_{cm} - x''_{cm})^2/4b^2}e^{-(x''_{cm} - x'_{cm})^2/4b^2}\ket{x_{rel}}\bra{x'_{rel}}\\
    &&\hspace{2cm}+ e^{-(x'_{cm} - x''_{cm})^2/4b^2}e^{-(x''_{cm} - x_{cm})^2/4b^2}\ket{x'_{rel}}\bra{x_{rel}}\left.+ e^{-(x''_{cm} - x'_{cm})^2/2b^2}\ket{x'_{rel}}\bra{x'_{rel}}\right].
\end{eqnarray}
After integrating the probability becomes,
\begin{eqnarray}\nonumber
    p(x_{cm}) &=& \frac{1}{b\sqrt{2\pi}}\left[\left(\erf\left(\frac{x_{cm}-x_{cm,i}}{b\sqrt{2}}\right) - \erf\left(\frac{x_{cm}-x_{cm,i}-\Delta x_{cm}}{b\sqrt{2}}\right)\right)\ket{x_{rel}}\bra{x_{rel}}\right.\\\nonumber
    &&+ e^{-(x_{cm}-x'_{cm})^2/8b^2}\left(\erf\left(\frac{x_{cm}+x'_{cm}-2x_{cm,i}}{b\sqrt{8}}\right)\right.\\\nonumber
    &&- \left.\erf\left(\frac{x_{cm}+x'_{cm}-2(x_{cm,i}+\Delta x_{cm})}{b\sqrt{8}}\right)\right)\times(\ket{x_{rel}}\bra{x'_{rel}} + \ket{x'_{rel}}\bra{x_{rel}})\\
    &&+ \left.\left(\erf\left(\frac{x'_{cm}-x_{cm,i}}{b\sqrt{2}}\right) - \erf\left(\frac{x'_{cm}-x_{cm,i}-\Delta x_{cm}}{b\sqrt{2}}\right)\right)\ket{x'_{rel}}\bra{x'_{rel}}\right],\label{eqn:ZmodelProb}
\end{eqnarray}
where $\erf$ denotes the Error function. When the position states become highly localized, \emph{i.e.} $b\to 0$, the uncertainty of the detector becomes small,\emph{i.e.} $\Delta x\to0$. In this limit (\ref{eqn:ZmodelProb}) becomes,
\begin{eqnarray}\nonumber
    p(x_{cm}) &=& \frac{1}{b}\left[\Theta(x_{cm}-x_{cm,i}) - \Theta(x_{cm}-x_{cm,i}-\Delta x)\right]\ket{x_{rel}}\bra{x_{rel}}\\ 
    &+& \frac{1}{b}\left[\Theta(x'_{cm}-x_{cm,i}) - \Theta(x'_{cm}-x_{cm,i}-\Delta x)\right]\ket{x'_{rel}}\bra{x'_{rel}}.\label{ZmodelProbResult}
\end{eqnarray}
The cross-terms have vanished since the exponential becomes a delta function in the highly localized limit and $x_{cm}\neq x'_{cm}$. When the measured center-of-mass position is within a bin of the detector one of the coefficients of (\ref{ZmodelProbResult}) is nonzero. As the energy of the system is increased the measured result will be close to $x_{cm,i}$, \emph{i.e.} $\Delta x\to0$. This means that the $\Theta$-functions will become closer making the entire coefficient infinitesimally narrow and infinitely tall. The end result will be a mixed state who's density matrix is $\rho = \ket{x_{rel}}\bra{x_{rel}} + \ket{x'_{rel}}\bra{x'_{rel}}$, matching the results from the $Z$-model.
\end{widetext}
%----------------------\\

\section{Conclusions}\label{sec:Conclusion}
In this paper we have implemented localization of quantum information as an entanglement extraction protocol in relational systems with Gaussian states.  As one might expect, when full translation invariance is implemented via the $G$-twirl, leaving only relational states, localization can't matter and there should be no notion of an entanglement extraction process.  We find that this is indeed the case, as the $G$-twirl not only wipes out information on the external frame, it also naturally wipes out any entanglement amongst center-of-mass and relational degrees of freedom that were present in the external partition.  The resultant mixed state has no entanglement to extract.

In contrast, if one keeps the external partition information but implements it relationally via a $Z$-model, then there is entanglement that can be extracted and states can be localized.  In this scenario, however, the implementation of the $Z$-model itself creates a Hamiltonian in which there is an energy difference between the initial (entangled) and final (localized, unentangled) states.  Hence entanglement extraction procedures as outlined in Hackl can be run as expected.  We expect that a similar outcome would be present if one used different otherwise conserved quantities than the energy to label states\textemdash entanglement extraction would always require some net gain or loss of some resource (for example charge or angular momentum).  We further found there is a smooth map between the two relational constructions, which can be implemented in the language of POVM's.

Our work broke the translation invariance and hence the degeneracy of the Hamiltonian via a simple dynamical method we imposed by hand, that of a $Z$-model, or external field.  However, gravitational self-interactions would also in principle lead to a non-degenerate Hamiltonian for the initial and final states.  In this type of scenario, the extraction and localization process would result in a) the amount of entanglement extracted being proportional to the gravitational self-energy difference of the two states, and simultaneously b) the system becoming more localized in space. This, as one might expect, qualitatively reflects black hole thermodynamics and other holographic approaches connecting entanglement and gravitational dynamics.  Whether or not implementing such a localization/extraction protocol in this framework quantitatively matches black hole physics we leave for future work.

\acknowledgments
A. Dukehart and D. Mattingly thank the Department of Energy for support through DOE grant DE-SC0020220.  We also thank A. Smith for useful comments on a draft of this manuscript.

\bibliography{LocalizationofRelationalQI}

\end{document}